\documentclass[12pt]{article}

\usepackage{amssymb,amsmath}

\usepackage{graphicx}
\DeclareGraphicsExtensions{.pdf,.png,.jpg}

\newcommand{\beq}[1]{\begin{equation}\label{#1}}
\newcommand{\eeq}{\end{equation}}
\newcommand{\bear}[1]{\begin{eqnarray}\label{#1}}
\newcommand{\ear}{\end{eqnarray}}

\newcommand{\p}{\partial}
\newcommand{\tr}{ {\rm tr} }

\newcommand{\fnm}{\footnotemark}
\newcommand{\fnt}{\footnotetext}

\newcommand{\eps}{ \varepsilon }

\newcommand{\R}{ {\mathbb R} }

 \catcode`\@=11 \@addtoreset{equation}{section}\catcode`\@=12

\begin{document}

 \begin{center}
 \large \bf
  Exact    $(1 + 3 + 6)$-dimensional cosmological-type  solutions in gravitational  model with Yang-Mills field,
  Gauss-Bonnet term and $\Lambda$-term

 \end{center}

 \vspace{0.3truecm}

 \begin{center}

 \normalsize\bf

 \vspace{0.3truecm}

   V. D. Ivashchuk$^{1,2}$, K. K. Ernazarov$^{1}$, and  A. A. Kobtsev$^{3}$

\vspace{0.3truecm}
  
  \it $^{1}$ 
    Institute of Gravitation and Cosmology, \\
    Peoples' Friendship University of Russia (RUDN University), \\
    6 Miklukho-Maklaya Street,  Moscow, 117198, Russian Federation, \\ 
 
 \vspace{0.1truecm}
 
  \it $^{2}$ Center for Gravitation and Fundamental Metrology,  VNIIMS, \\
  46 Ozyornaya Street, Moscow, 119361,  Russian Federation.
\vspace{0.1truecm}
 
 \it $^{3}$ Institute for Nuclear Research of the Russian Academy of Sciences,
  Moscow, Troitsk, 142190,  Russian Federation

 \end{center}

\begin{abstract}

We consider $10$-dimensional gravitational  model with $SO(6)$ Yang-Mills field,
  Gauss-Bonnet term and $\Lambda$-term.
 We  study so-called cosmological type solutions  defined on product manifold 
 $M = \R \times \R^3 \times K$, where 
 $K$  is $6d$  Calabi-Yau manifold. By putting the gauge field 1-form  
 to be coinciding with  1-form  spin connection on $K$, 
we  obtain exact  cosmological  solutions with exponential dependence of scale factors
(upon $t$-variable), governed by two non-coinciding Hubble-like parameters: $H >0$, $h$, 
obeying $ H + 2 h \neq 0$. We also present static analogs of these cosmological solutions
(for $H \neq 0$, $h \neq H$ and $ H + 2 h \neq 0$). 
The islands of stability for both classes of solutions are outlined.

\end{abstract}

\vspace{0.1truecm}


\vspace{0.1truecm}

\section{Introduction}

Here we deal with a so-called  Einstein-Gauss-Bonnet-Yang-Mills-$\Lambda$  gravitational model in dimension  $D = 10$. 
The action of the model  contains scalar curvature,  Gauss-Bonnet, cosmological term ($\Lambda$-term) and Yang-Mills term with a value in $so(6)$ Lie algebra. The model  includes  non-zero constant $\alpha$, coupled to the sum of Yang-Mills and  Gauss-Bonnet terms. The equations of motion for this model are of  second order (like it takes place  in General Relativity). The so-called Gauss-Bonnet term  has appeared in (super)string theory  as a second order correction in curvature  to the  effective (super)string effective action   \cite{Zwiebach,FrTs2,GW} for a heterotic string \cite{GHMR}.  
 
 At present, Einstein-Gauss-Bonnet (EGB) gravitational models, e.g. with cosmological term and extra matter fields and its modifications   \cite{Ishihara}-\cite{Ivas-20},  are under intensive  studyies in astrophysics and cosmology. The main goal in these studies is a solution of dark energy problem. One can study such models for  possible explanation  of  accelerating  expansion of the Universe, which was supported by supernovae (type Ia) observational data \cite{Riess,Perl}. 
 
  We note that at present there exist several modifications of Einstein and EGB actions which correspond to $F(R)$, $R + f({\cal G})$, $f({R, \cal G})$, $f({R, \cal G})$ , $f({R, \cal G},T_{..}T^{..})$   theories (e.g. for $D=4$), where $R$ is scalar curvature and  ${\cal G}$ is  Gauss-Bonnet term. These modifications are under intensive studying  devoted to cosmological, astrophysical and other applications, see \cite{NOO-17}-\cite{HRF} and references therein. 

Another point of interest is a search of possible local manifestation of dark energy related to wormholes, black holes,  etc.  The  most  important results for black holes in models
with  Gauss-Bonnet term  are related  with  the  Boulware-Deser-Wheeler solution   \cite{BoulDes,Wheel} and its generalizations  \cite{Wheel2,Wilt,Cai,CvetNojOd},  see also Refs. \cite{GarGir,Charm,ABK} and references therein.
 For certain applications of  brane-world models with  Gauss-Bonnet term, see Refs.
 \cite{BrKonMel,TABLM} and related bibliography. For wormholes solutions in 
Einstein-Gauss-Bonnet models with certain fields see Refs. \cite{KKK,BKKK} and references threin.

In this article  we  deal with the so-called cosmological type solutions with the $10d$ metric  
   \begin{equation}
     g^{(10)} = - w d \chi \otimes d \chi  + a^2_3(\chi )g^{(3)} + a^2_6(\chi )g^{(6)}. \label{0.1}  
    \end{equation} 
defined on product manifold $\R \times \R^3 \times K,$   
 where $w = \pm 1$, $\R^3$ is  flat $3d$ manifold (``our'' space) with the metric $g^{(3)}$ and  $K$  is $6d$ Ricci-flat Calabi-Yau manifold (internal space) of $SU(3)$ holonomy group   with the metric $g^{(6)}$. The warped product model is governed by two scale factors, depending upon one variable $\chi $. It is the synchronous time variable for cosmological case $w =  1$:
$\chi  = t$, while it  coincides with space-like variable for  $w = -1$: $\chi  = u$.
The presence of Yang-Mills field  makes this ansatz  consistent if we choose the Lie algebra for Yang-Mills field to be equal (at least) to $so(6)$, which contains   $su(3)$  subalgebra, corresponding to $SU(3)$ group of golonomy of $6d$ Calabi-Yau manifold. For Yang-Mills field we consider the following ansatz:  we put here the gauge field 1-form to be equal to  spin connection 1-form  on $K$  (see Section 2):  $A =  \omega^{(6)}$. 
In such ansatz the gauge field plays a role of compensator which ``waves out'' the terms with non-zero Riemann tensor of Calabi-Yau metric $g^{(6)}$.  

Originally such idea of compensation was used by Wu and Wang \cite{WW}  (see also \cite{IM-14}) in $(1+3+6)$-dimensional cosmological model based on $10d$ Yang-Mills ($SO(32)$- and/or $E_8 \times E_8$-) supergravity theory  ``upgrated'' by additions of Chern-Simons and Gauss-Bonnet terms (of superstring origin).   The work of Wu and Wang was influenced greatly by the well-known paper of Candelas et al. \cite{CHSW}   devoted to  vacuum configurations in ten-dimensional $O(32)$ and $E_8 \times E_8$ supergravity and superstring theory that have unbroken $N = 1$ supersymmetry in four dimensions. 

It should be noted that  compactifications of  $11$-dimensional supergravity on ($6d$)
Calabi-Yau manifold were considered in Refs.   \cite{Duff,CCAF}. Moreover, $4d$ and $6d$ Calabi-Yau
manifolds also appeared in  partially sypersymmetric solutions of $D=11$ supergravity  with $M$-branes, see  Refs. \cite{DLPS,Gol-Ivas,Ivas-16s} and and references threin.

In Section 3 we obtain exact cosmological  solutions  with exponential dependence of scale factors (upon $t$-variable), governed by two non-coinciding Hubble-like parameters: $H >0$ and $h$, corresponding to factor spaces of dimensions $3$  and $6$, respectively, when  the  following restriction $ 3H + 6 h \neq 0$ is used (excluding the solutions with constant volume factor). 

  In Section 4 we  obtain static solutions ($w = - 1$) for non-coinciding Hubble parameters  $H \neq 0$, $h$, which obey  $ 3H + 6 h \neq 0$.  We also study  stability (in certain restricted sence) of the obtained  solutions in cosmological case for $t \to  + \infty$ (see Section 3)  and in static case for $u \to \pm \infty$ (see Section 4) by using results of Ref. \cite{Ivas-16} (see also  approach of Ref. \cite{Pavl-15}) and  single out  the subclasses of stable/non-stable solutions.

\section{The $10$-dimensional  model}

\subsection{The action and equations of motion}

 We take the action of the model as 

\begin{eqnarray}
S = \frac{1}{2 \kappa^2} \int d^{10}x\sqrt{-g} \left\{ R 
  + \alpha \left[  \tr F_{MN} F^{MN}  
  \right.  \right.  \nonumber    \\ \left. \left.
    + R_{MNPQ}R^{MNPQ} -4 R_{MN} R^{MN}  + R^2  \right]   \right\}, \label{1} 
\end{eqnarray}
where  $\kappa^2$ is  $10$-dimensional gravitational constant,
 $\alpha \neq 0$ is  constant, $g_{MN}$ are components of the metric, 
 $F_{MN}$ are components
of the Yang-Mills field strengths corresponding to $2$-form with the value in the Lie algebrs
$so(6)$:  $ F = \frac{1}{2} F_{MN}dx^M\wedge dx^N = dA + A \wedge A,$ 
where $A=A_Mdx^M$ is the $1$-form with the value in $so(6)$ 
($F_{MN} = \p_M A_N - \p_N A_M +[A_M,A_N]$).

 The action (\ref{1})  leads us to the following equations of motions:
\begin{eqnarray}
 R_{MN}-\frac{1}{2}g_{MN}R =     - 2 \alpha \biggl( \tr
 F_{MP}F_N^{\ \ P}-\frac{1}{4}g_{MN}\ \tr F_{PQ} F^{PQ} \biggr)
        \qquad  \qquad     \nonumber \\
     + \alpha \biggl[ \frac{1}{2}g_{MN}( R_{MNPQ}R^{MNPQ} -4 R_{MN} R^{MN}  + R^2 )
 - 2RR_{MN}   \nonumber \\
 + 4R_{MP}R_N^{\ \ P} + 4R_{MPNQ}R^{PQ} - 2R_M^{\ \ PQS}R_{NPQS} \biggr],   \label{4}
\end{eqnarray}

\begin{equation}
  D_M F^{MP} = 0. \label{6}  
  \end{equation}
Here  we use notation for covariant gauge derivative: $D_M=D_M(A)=\nabla_M+[A_M,.]$.

\subsection{Cosmological ansatz. }

Let us consider  ten-dimensional manifold
\begin{equation}
   M = \R \times \R^3 \times K ,    \label{8}
\end{equation}
where $K$ is a $6d$ Calabi-Yau manifold, i.e., a compact 6-dimensional K\"ahler
Ricci-flat manifold with the metric, which has $SU(3)$ holonomy group. 
For corresponding Lie algebra we have $su(3) \subset so(6)$.

We start with the cosmological case, i.e.
we consider the set of equations (\ref{4}), (\ref{6}) 
on the manifold (\ref{8}) with  the following ansatz for  fields
\begin{eqnarray}
   g^{(10)} = -dt\otimes dt + a^2_3(t)g^{(3)} + a^2_6(t)g^{(6)},
   \label{10} \\
   A =  \omega^{(6)}.
 \label{13}
 \end{eqnarray}

Here 
$g^{(3)} =  dx^1 \otimes dx^1 + dx^2 \otimes dx^2 + dx^3 \otimes dx^3$, i.e. 
we deal with a flat Euclidean metric on $\R^3$  and   
$g^{(6)} = g^{(6)}_{mn}(y) dy^m \otimes dy^n $  is  Calabi-Yau metric on  $K$.  

By $\omega^{(6)} = \omega^{(6)}_{m}(y) dy^{m} $ we denote the spin connection 1-form on $K$  with 
the value in the Lie algebra $so(6)$ 
\fnm[1]\fnt[1]{In fact it belongs to subalgebra  $su(3) \subset so(6)$.} corresponding to the
(local) basis of co-vectors $e^{a}_{\ m}$ on $K$ , which diagonalizes the metric $g^{(6)}$: 
 \begin{equation}
 g^{(6)}_{m n }  = e^a_{\ m} e^b_{\ n} \delta_{ab},    \label{10e}
  \end{equation}
 \begin{equation}
  \omega^{(6)}_{m}  = \mid \mid \omega^{(6) a}_{\ \ \ \ b m}\mid \mid \equiv \mid \mid e^a_{\ n}
                  \nabla^{(6)}_{m} e^{n}_{\ b} \mid \mid \subset so(6),    \label{10s}
   \end{equation}      
where covariant derivative $\nabla^{(6)}_{m}$ corresponds to the metric 
$g^{(6)}$  and  $e^{n}_{\ b}$ is ``inverse'' (dual) basis of vector fields obeying
  $e^a_{\ n} e^{n}_{\ b} = \delta^{a}_{b}$.

 It follows from (\ref{13}) that
 \begin{equation}
 F = \Omega^{(6)} ,  \label{13a}
 \end{equation}
 where $\Omega^{(6)}= d \omega^{(6)} + \omega^{(6)} \wedge \omega^{(6)} $ is the curvature two-form
 on $K$ with the value in  $so(6)$.

The spin connection $\omega^{(6)}$ on $K$ obeys the identity
\begin{equation}
 D_m(\omega^{(6)})\Omega^{(6) mn} = 0, \label{18}
\end{equation}
where $D_m(\omega^{(6)})=\nabla^{(6)}_m + [\omega^{(6)}_{m},. \ ]$. 
The identity (\ref{18}) is equivalent to the following identity for the Riemann tensor on $K$
\begin{equation}
 \nabla^{(6)}_m R^{(6)mnpq} = 0, \label{18i}
\end{equation}
which is valid for any K\"ahler Ricci-flat manifold \cite{WitW}.

The Yang-Mills equations (\ref{6}) are satisfied identically due to 
eqs. (\ref{10}), (\ref{13}) and (\ref{18}).

Let us denote
\begin{equation}
 H(t) \equiv \dot{a}_3/a_3, \qquad   h(t) \equiv \dot{a}_6/a_6,                 \label{19b}
\end{equation}
where  in this section we denote $\dot{a} \equiv \frac{d a}{dt}$.

Then, equations (\ref{4}) for the ansatz
(\ref{10})-(\ref{13}) may be written as follows:
\begin{eqnarray}
 B_0 + 2 \Lambda  - \alpha B_1 =0, 
    \label{20a} \\
 \frac{d L_H}{dt}  +  (3H + 6h) L_H -  L_0 = 0,  \label{20b}  \\
 \frac{d L_h}{dt}  +  (3H + 6h) L_h -  L_0 = 0.   \label{20c}
     \end{eqnarray}

Here 
      \begin{eqnarray}
      L_0 =  B_0 - 2 \Lambda - \frac{1}{3} \alpha B_1,
      \label{21.L0} \\                
       L_H =  2  B_H  - \frac{4}{3} \alpha A_H  
       \label{21.H}, \\
       L_h =  2  B_h  - \frac{4}{3} \alpha A_h,  \label{21.h}   
       \end{eqnarray}
       
       \begin{eqnarray}
            B_0 = 3H^2 + 6h^2 - (3H + 6h)^2, \label{22.B0} \\ 
            \quad B_H = H - (3H + 6h) \label{22.BH}, \\
            \quad B_h = h - (3H + 6h),  \label{22.Bh}            
       \end{eqnarray}
                
        \begin{eqnarray}
           B_1 =  144 H^3 h + 1080 H^2 h^2 + 1440 H h^3  + 360 h^4
                \label{23.B1}
         \end{eqnarray}
           and
             \begin{eqnarray}
              A_{H} = 36 H^2 h  + 180 H h^2 + 120 h^3,
                                 \label{23.AH} \\
              A_{h} = 6 H^3 + 90 H^2 h + 180 H h^2 + 60 h^3 .
                   \label{23.Ah}
                 \end{eqnarray}

Equations (\ref{20a})-(\ref{20c})  are obtained from
(\ref{4}) using the Ricci flatness of
$K$ and the equality for Riemann tensor  of internal space $K$ with the metric $g^{(6)}$

\begin{equation}
 R^{(6)}_{pqmn} R^{(6)qp'mn} =  g^{(6)mp'}g^{(6)nq} \ {\rm tr} \ F_{mn}F_{pq},             \label{19i}
\end{equation}

which follows from (\ref{13a}) and  well-known identity

\begin{equation}
  R^{(6)p}_{\ \ \ \ qmn} = e^{p}_{\ a} \
 e^{b}_{\ q} \ \Omega^{(6) a}_{\ \ \ \ b mn} \ .             \label{19ii}
\end{equation}

\section{Cosmological solutions }

 Here we consider the case when Hubble-like parameters are
 constant, i.e. 
 \begin{equation}
  H(t) = H = {\rm const}, \qquad   h(t) = h = {\rm const}.                \label{19}
 \end{equation}
 
 For scale factors we get the exponential dependence on $t$  
 \begin{equation}
   a_3(t) = \exp(H t), \qquad   a_6(t) = \exp(h t).                \label{19a}
  \end{equation}

 We get a set of polynomial equations
  \begin{eqnarray}
   B_0 + 2 \Lambda  - \alpha B_1 =0, 
      \label{24a} \\
   (3H + 6h) L_H -  L_0 = 0,  \label{24b}  \\
    (3H + 6h) L_h -  L_0 = 0,   \label{24c}
       \end{eqnarray}
  where polynomials $B_0, B_1, L_0, L_H, L_h$ are defined above.

We set 
\begin{equation}
  \label{33.2a}
   H > 0. 
\end{equation}
This relation is used for a description of an accelerated expansion of the
$3$-dimensional subspace (which may describe our Universe).
The evolution of the $6$-dimensional internal factor space is described  by the Hubble-like  parameter $h$.

It follows from Ref. \cite{Ivas-16,IvKob-18-2a} (for a more general splitting scheme
see paper by Chirkov, Pavluchenko and Toporensky  \cite{ChPavTop1})
 that if we consider  Hubble-like parameters $H$ and $h$  
obeying two restrictions imposed
   \begin{equation}
   3 H + 6h \neq 0, \qquad  H \neq h,
   \label{33.3}
   \end{equation}
 we reduce relations (\ref{24a}), (\ref{24b}), (\ref{24c})
  to the  following set of  equations

\begin{eqnarray}
%\begin{aligned}
E =  3 H^2 + 6 h^2 - (3H + 6h)^2  + 2 \Lambda \qquad
\nonumber \\
- \alpha [ 144 H^3 h         
+ 1080 H^2 h^2 + 1440 H h^3  + 360 h^4] = 0,
\label{33.4}
%\end{aligned}
\end{eqnarray}

\begin{equation}
\label{33.5}
  Q =   2 H^2 + 20 H h    + 20 h^2 = - \frac{1}{2 \alpha}.   
\end{equation}

Using equation  (\ref{33.5}), we get 
\begin{equation}
H   =     (- 2 \alpha {\cal P})^{-1/2}, 
 \label{33.6}
\end{equation}
where 

\begin{eqnarray}
{\cal P}  =  {\cal P}(x)  \equiv    2 + 20 x  + 20 x^2, 
 \label{33.7}  \\
    x  \equiv h/H,
    \label{33.7x}
 \end{eqnarray}
and 
 \begin{equation}
\alpha {\cal P} < 0. 
 \label{33.8} 
\end{equation}

Due to restrictions (\ref{33.3})  we have for $x$ from (\ref{33.7x})
\begin{equation}
  x \neq x_d  \equiv  - 1/2, \qquad  x \neq x_a \equiv 1. 
 \label{33.8da} 
\end{equation}
 
 The relation  (\ref{33.5}) is valid only if        
\begin{equation}
  {\cal P}(x) \neq 0.  \label{33.8b}
\end{equation}
%For ${\cal P}(x) = 0$ the equation (\ref{33.5}) is not satisfied.

Substituting  relation (\ref{33.6}) into (\ref{33.4})
we obtain  
\begin{eqnarray}
    \Lambda \alpha \equiv  \lambda  = f(x)  \equiv
     (1/4)[3 + 6x^2 - (3 + 6x)^2]( 2   + 20x  + 20x^2 )^{-1} \nonumber \\
   + (1/8)[18(8x    + 60x^2 + 80x^3    + 20x^4)]( 2   + 20x  + 20x^2 )^{-2} . 
    \label{33.8L}
    \end{eqnarray}

From (\ref{33.8b}) we get
 \begin{equation} 
  x \neq x_{\pm}  \equiv \frac{- 10 \pm \sqrt{10}}{20}, 
 \label{3.9}
 \end{equation}
where $x_{\pm}$ are roots of the quadratic equation ${\cal P}(x) = 0$.
They obey 
\begin{equation} 
    x_{-} < x_{+} < 0. 
         \qquad \label{33.11} 
  \end{equation}

According to eq. (\ref{33.8}) we get (in this cosmological case)
\begin{equation} 
    x_{-} < x <  x_{+} \ \ {\rm for} \  \alpha > 0,  
               \qquad \label{33.11a}
 \end{equation}    
  and    
   \begin{equation}  
     x <  x_{-} \ {\rm or} \ x > x_{+} \ \ {\rm for} \  \alpha < 0.
                 \qquad \label{33.11b}
 \end{equation}

  The graphical representation of the  function $\lambda = f(x)$ is given at Figure 1. 
   
   The function $f(x)$ obeys 
    \begin{equation}  
     \lim_{x \to \pm \infty} f(x) = 
      \lambda_{\infty} \equiv  - \frac{21}{80}
                     \qquad \label{33.8xinf}
   \end{equation}
  and 
  \begin{equation}  
     \lim_{x \to x_{\pm} } f(x) = - \infty.
                     \qquad \label{33.8xpm}
   \end{equation}
  It has a    maximum at 
  $x_c = - \frac{1}{5}$ with the value $f(x_c) = \lambda_c \equiv  3/10$
  and an inflection point
  at $x_d = - \frac{1}{2}$ with the value $f(x_d) = \lambda_d \equiv  3/16$.
  It has also a point of local maximum 
  at $x_a \equiv 1$ with the
  value $f(x_a) = \lambda_a \equiv - 3/14$. 
   
  % By using the relation for the function $$

 Equation (\ref{33.8L}) is equivalent to the following master
 equation
 \begin{equation} 
    (400 \lambda +105) x^4 + (800 \lambda +150) x^3 + 
    (480 \lambda+90) x^2 + (80 \lambda+30)x+ 4\lambda +3 = 0. 
           \label{33.M} 
   \end{equation}

For 
 \begin{equation} 
 \lambda  \neq \lambda_{\infty} \equiv -  21/80 \label{33.noninf} 
 \end{equation}
the solution to this (fourth order) master equation reads
\begin{eqnarray}
x = \eps_1 \frac{1}{2} \sqrt{ - \frac{\eps_2 A}{\sqrt{X}} - Y^{1/3} + B Y^{-1/3} + C}
\nonumber  \\
 - \eps_2 \frac{\sqrt{X}}{2 \sqrt{5}(80 \lambda  +21)}  
 - \frac{5(16 \lambda +3)}{2(80 \lambda+21)}, \label{30x}
\end{eqnarray}                                                                      
where $\eps_1 = \pm 1$,   $\eps_2 = \pm 1$, 

\begin{eqnarray}
 A =  (18 \sqrt{5}(1280 \lambda^2+96 \lambda -39))/(80 \lambda + 21)^2,
 \label{31A}  \\
B =  -(4(16 \lambda-3)(20 \lambda+3))/(5(80 \lambda + 21)^2),
\label{31B}  \\
C =    (2(16 \lambda+3)(80\lambda-9))/(80\lambda +21)^2,
\label{31C} 
\end{eqnarray}
and
\begin{eqnarray} 
X = 5(80 \lambda+21)^2 Y^{1/3} +  5(16 \lambda +3)(80 \lambda-9)  +  4(16 \lambda-3)(20 \lambda+3)Y^{-1/3}, 
  \label{32X} \\ 
 Y = (36(16 \lambda -3) \sqrt{-(10 \lambda-3)(14 \lambda+3)})/(5^{3/2}(80 \lambda + 21)^3) + Z,
 \label{32Y} \\ 
 Z = (4(16 \lambda -3)(320 \lambda^2+42 \lambda-9))/(5(80 \lambda+21)^3).
  \label{32Z}  
\end{eqnarray}

\begin{figure}[!h]
 	%\begin{minipage}[h]{0.49\linewidth}
 		\begin{center}
 			\includegraphics[width=\linewidth]{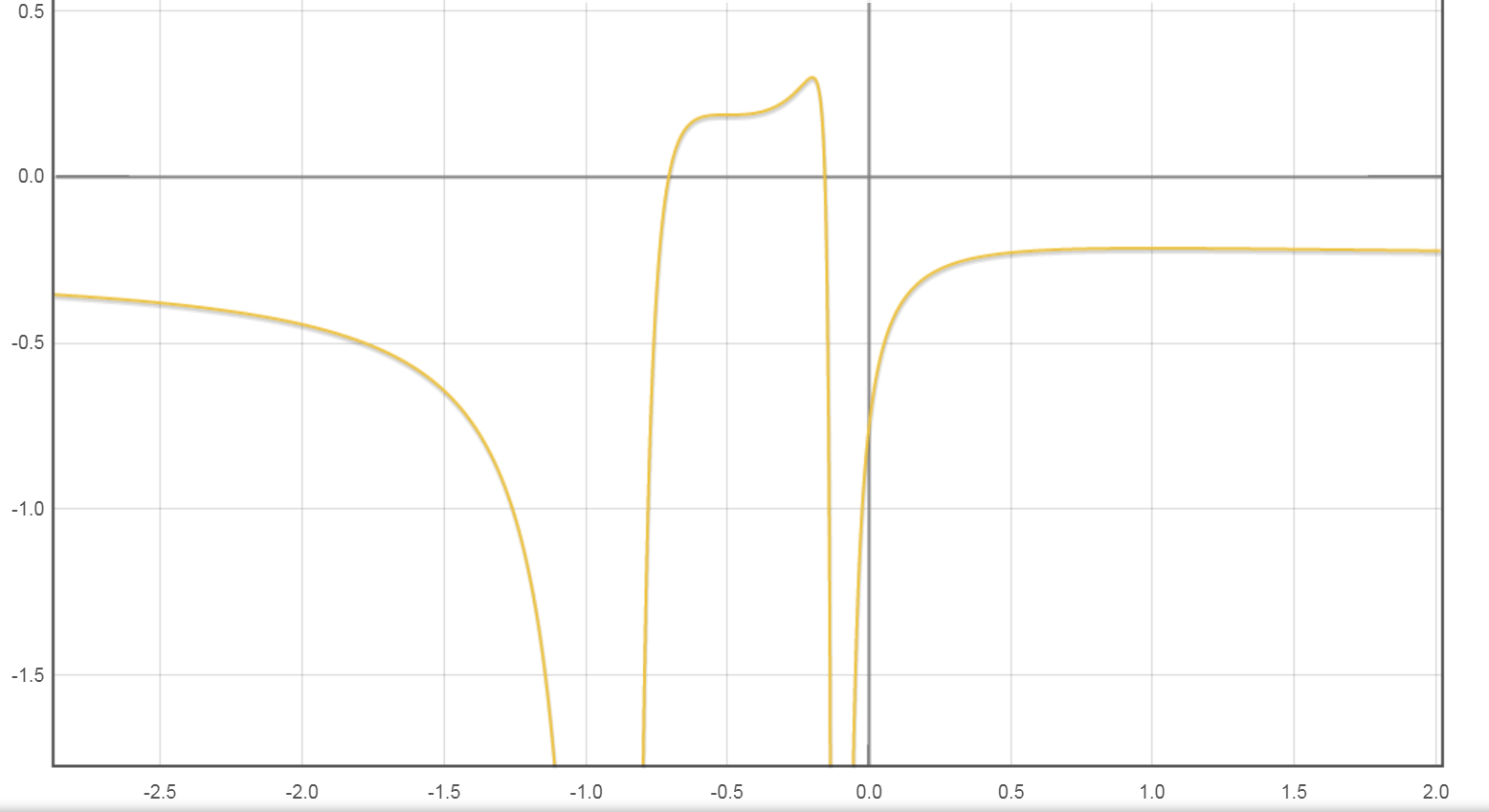}
 			\caption{The graphical representation of moduli function $\lambda = f(x)$  given by relation (\ref{33.8L}).}
 			\label{fig:1}
 		\end{center}
 	%\end{minipage}
  \end{figure}

It follows from Figure 1 that for given parameters  $\Lambda$ and $\alpha$, obeying restrictions $\alpha \neq 0$
and  (\ref{33.noninf}), real solutions in formula (\ref{30x}) appear 
for suitably chosen $\eps_1 = \pm 1$ and  $\eps_2 = \pm 1$ if 
\begin{equation}
 \lambda = \Lambda \alpha   \leq \lambda_c = 3/10 
     \label{32c}
     \end{equation}                                                                  
for $\alpha > 0$ and 
\begin{equation}
 \lambda = \Lambda \alpha   < \lambda_a =  - 3/14
     \label{32a}
     \end{equation} 
for $\alpha < 0$. 

For example, for $\lambda = \lambda_c = 3/10 $ relation (\ref{30x}) gives us $x = x_c = -1/5$ 
if we put $\eps_1 = \pm 1$ and  $\eps_2 = - 1$.

In exceptional case  
 \begin{equation} 
 \lambda  = \lambda_{\infty} = -  21/80 \label{33.inf} 
 \end{equation}
we have a cubic master equation  (\ref{33.M}) which has three real roots
\begin{equation} 
x = x_k = (1/10)(- 2 -  6 \cos{(\theta/3 + 2k \pi/3) }), \qquad \tan{\theta}  = \sqrt{15}
\label{32Xex},
\end{equation}
$k = 1,2,3$, or, numerically
\begin{equation}
x_1 \approx 0.29253505115, \quad x_2 \approx -0.14952367803, \quad  x_3 \approx -0.74301143583.
\label{32Xnum}
\end{equation}

{\bf Graphical analysis.}
The graphical representation  $\Lambda |\alpha|$  upon $x = h/H$ (in this cosmological case)
is presented at Figure 2. In drawing this figure   
we use the relation
\begin{equation}
  \Lambda |\alpha|  = \lambda {\rm sgn}(\alpha) = 
  f(x) \frac{(- 2 - 20x - 20x^2)}{|2   + 20x  + 20x^2|}, 
     \label{32Lc}
    \end{equation} 
in agreement with  (\ref{33.7}) and (\ref{33.8}). It follows from Figure 2 that
real solutions take place if
\begin{equation}
  \Lambda |\alpha|  \leq \lambda_c = 3/10 , 
     \label{32pc}
    \end{equation} 
for $\alpha > 0$ and
\begin{equation}
  \Lambda |\alpha|  > |\lambda_a| = 3/14, 
     \label{32pa}
    \end{equation}  
for $\alpha < 0$. (The point $x_a = 1$ is excluded from our consideration.)

\begin{figure}[!h]
 	%\begin{minipage}[h]{0.49\linewidth}
 		\begin{center}
 			\includegraphics[width=\linewidth]{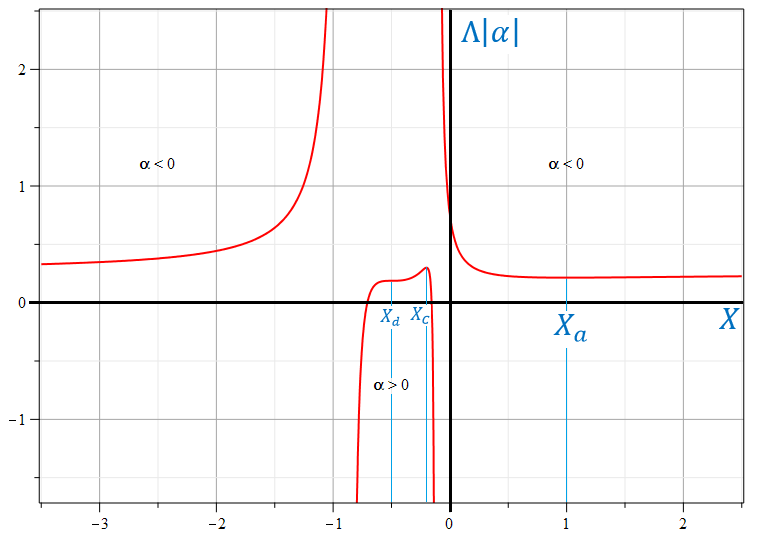}
 			\caption{The dependence of $\Lambda |\alpha|$  upon $x = h/H$ in cosmological case.
 			The central branch coresponds to $\alpha >0$. The left and right branches corespond to $\alpha < 0$. }
 			\label{fig:2}
 		\end{center}
 	%\end{minipage}
  \end{figure}

{\bf Stability.}

Using results of Ref. \cite{Ivas-16,IvKob-18-2a} we obtain that
cosmological solutions under consideration obeying  $x = h/H \neq x_i$, $i = a,c,d$,
where $x_a =1$,  $x_c = - \frac{1}{5}$,  $x_d = - \frac{1}{2}$, 
are  stable if i) $x > x_d = -1/2$ and unstable if ii) $x < x_d = -1/2$.
\fnm[2]\fnt[2]{For isotropic cosmological solutions with $H=h$, 
see Refs.  \cite{ChPavTop,Ivas-16} for 
generic $\Lambda$ and  \cite{Iv-09,Iv-10} for $\Lambda = 0$.} 

We note that  the   the points $x_a =1$ and $x_d = -1/2$ 
are excluded from our consideration due to restrictions (\ref{33.3})
while the point of maximum $x_c = - \frac{1}{5}$ is excluded 
since the analysis of Ref. \cite{Ivas-16} was based on  
the equations for perturbations for $\delta H(t)$, $\delta h(t)$
in the linear approximation which can be resolved when 
$x_c \neq - \frac{1}{5}$. In special case $x_c = - \frac{1}{5}$ 
higher order terms in pertubations should be considered.  

Let us denote by  $n_{s}$ the number of non-special stable solutions. 
By using Figure 2 we find just graphically for $\alpha > 0$

   \begin{equation}
     n_{s} = \begin{cases}
     0, \  \Lambda \alpha \geq \lambda_c = 3/10, \\
     2, \ \lambda_d = 3/16 < \Lambda \alpha < \lambda_c = 3/10, \\
     1, \  \Lambda \alpha \leq \lambda_d = 3/16. \\
    \end{cases}   \label{40s}
   \end{equation}
 (here $\lambda_i= f(x_i)$) while    for  $\alpha < 0$ we obtain          
    \begin{equation}
      n_{s} = \begin{cases}
      1, \  \Lambda |\alpha| \geq |\lambda_{\infty}| = 21/80, \\
      2, \  |\lambda_{a}| = 3/14 < \Lambda |\alpha| < |\lambda_{\infty}| = 21/80, \\
      0, \   \Lambda |\alpha| \leq |\lambda_{a}| = 3/14. \\
     \end{cases}   \label{41s}
    \end{equation}
  Thus,  for $\alpha > 0$ and small enough value of $\Lambda$ there exists
   at least one stable solution  with $x \in (x_{-},x_{+})$, while 
   for $\alpha < 0$ and big enough value of $\Lambda$
 there exists at least one stable solution with $x$ obeying $x > x_{+}$. The solutions with $x < x_{-}$  
 are unstable.
 
 In  cosmological case real solutions corresponding to $\Lambda = 0$ exists only if $\alpha > 0$.
 We obtain from (\ref{30x}) for $\eps_1 = \pm 1$,   $\eps_2 = 1 $, two solutions  
  \begin{eqnarray}
 x_{-} \approx -0.70692427923, \qquad      x_{+} \approx - 0.15626295995. 
 \end{eqnarray}                                                                      
The first cosmological solution (for $x_{-}$) is unstable, while the second one  (for $x_{+}$)
is stable in agreement with Ref. \cite{IvKob-19-EPJC}.

 {\bf Remark.} It should be noted that here as in the Ref. \cite{Ivas-16}  we 
 deal with restricted stability problem. We do not consider the general setup for
 perturbations $\delta g_{MN}(t,x)$ and $\delta A_{M}(t,x)$ but only consider the cosmological
 perturbations of scale factors $\delta a_3(t)$, $\delta a_6(t)$ in the framework of our ansatz
 (\ref{10}), (\ref{13}) with fixed $g^{(10)}$ and $\omega^{(6)}$. Analogous remark should
 be addressed to our analysis of static solutions in the next section.

 {\bf Zero variation of $G$.}
The cosmological solution with $x=0$, or $h = 0$, takes place if $\alpha < 0$ and
\begin{equation}
  \Lambda \alpha  = -  3/4. 
    \label{32var0} 
       \end{equation}
We get $\Lambda > 0$. The scale factor $a_6$ is constant in this case and we are led to zero variation 
of the effective $4d$ gravitational constant (in Jordan frame). This solution is stable.
Moreover, we get  $H^2 = - 1/(4 \alpha)$, which implies for the effective $4$-dimensional 
cosmological constant $\Lambda_{eff} = 3H^2 = \Lambda$. In general case 
$\Lambda_{eff}$ is a nontrivial function of $\Lambda$ and $\alpha$ given by (\ref{33.6})
and generic solution for $x$ from (\ref{30x}) (or from (\ref{32Xex}) in special case).

We note that for  $\alpha < 0$  and for  $\Lambda$ from (\ref{32var0}) there exists another
real solution corresponding to certain $x_{*} < x_{-}$, which is unstable.

\section{Static analogs of cosmological solutions }

Now with deal with the static case by 
 considering the set of equations (\ref{4}), (\ref{6}) 
on the manifold (\ref{8}) with  the following ansatz
\begin{eqnarray}
   g^{(10)} = du\otimes du + a^2_3(u)g^{(3)} + a^2_6(u)g^{(6)},
   \label{10u} \\
   A =  \omega^{(6)},
 \label{13u}
 \end{eqnarray}
where $u$ is a spatial coordinate and 
$g^{(3)} =  - dt \otimes dt + dx^1 \otimes dx^1 + dx^2 \otimes dx^2, $
is  flat  pseudo-Eucleadean metric on $\R^3$,    
$g^{(6)} $  is the Calabi-Yau metric on  $K$ and  
$\omega^{(6)}$ is spin connection 1-form on $K$
defined in a previous section.

The Yang-Mills equations  are satisfied identically as in the previous
case.

Now, we  denote
\begin{equation}
 H(u) \equiv \dot{a}_3/a_3, \qquad   h(u) \equiv \dot{a}_6/a_6,  \label{19bu}
\end{equation}
where  in this section we denote $\dot{a} \equiv \frac{d a}{du}$.

Then, equations (\ref{4}) in the ansatz
(\ref{10u})-(\ref{13u}) may be written as follows:
\begin{eqnarray}
 B_0  - 2 \Lambda  + \alpha B_1 =0, 
    \label{20au} \\
 \frac{d \bar{L}_H}{du}  +  (3H + 6h) \bar{L}_H  -  L_0 = 0,  \label{20bu}  \\
 \frac{d \bar{L}_h}{du} +  (3H + 6h) \bar{L}_h -  L_0 = 0,   \label{20cu}
     \end{eqnarray}

where 
      \begin{eqnarray}
      \bar{L}_0 =  B_0 + 2 \Lambda + \frac{1}{3} \alpha B_1,
      \label{21.L0u} \\                
       \bar{L}_H =  2  B_H  + \frac{4}{3} \alpha A_H  
       \label{21.Hu}, \\
       \bar{L}_h =  2  B_h  + \frac{4}{3} \alpha A_h,  
       \label{21.hu}   
       \end{eqnarray}
         $B_0, B_H, B_h, B_1$ are defined in  
        (\ref{22.B0}), (\ref{22.BH}), (\ref{22.Bh}), (\ref{23.B1}),
         and  $A_{H}, A_h$ are defined   in (\ref{23.AH}), (\ref{23.Ah}), respectively.
            
  As we see, the equations of motion for ``Hubble-like'' parameters (\ref{19bu}) in static case 
   may be obtained from cosmological ones (of Section 3) just by replacement
    \begin{equation}
     \alpha \mapsto - \alpha , \qquad   \Lambda  \mapsto - \Lambda. \label{21aL}
    \end{equation}
   The dimensionless parameter     $\lambda = \Lambda \alpha$ is invariant under 
   this replacement.   

Here we consider the case when ``Hubble-like'' parameters are
 constant, i.e. 
 \begin{equation}
  H(u) = H = {\rm const}, \qquad   h(u) = h = {\rm const},    \label{19u}
 \end{equation}
 or, equivalently, 
 \begin{equation}
   a_3(u) = \exp(H u), \qquad   a_6(u) = \exp(h u).        \label{19au}
  \end{equation}

 We get a set of polynomial equations
  \begin{eqnarray}
   B_0 - 2 \Lambda  + \alpha B_1 =0, 
      \label{24au} \\
   (3H + 6h) \bar{L}_H -   \bar{L}_0 = 0,  \label{24bu}  \\
    (3H + 6h) \bar{L}_h -  \bar{L}_0 = 0,   \label{24cu}
       \end{eqnarray}
  where polynomials $\bar{L}_0, \bar{L}_H, \bar{L}_h$ are defined by 
 relations (\ref{21.L0u}),  (\ref{21.Hu}), (\ref{21.hu}), respectively.

Here we consider slightly more general case   
\begin{equation}
  \label{33.2au}
   H \neq 0. 
\end{equation}

As in previous section we impose the conditions
(\ref{33.3}) and reduce relations (\ref{24au}), (\ref{24bu}), (\ref{24cu})
  to the  set of two equations \cite{Ivas-20}
\begin{eqnarray}
%\begin{aligned}
\bar{E} =  3 H^2 + 6 h^2 - (3H + 6h)^2  - 2 \Lambda \qquad
\nonumber \\
+ \alpha [ 144 H^3 h         
+ 1080 H^2 h^2 + 1440 H h^3  + 360 h^4] = 0,
\label{33.4u}
%\end{aligned}
\end{eqnarray}
 
\begin{equation}
\label{33.5u}
  Q =   2 H^2 + 20 H h    + 20 h^2 =  \frac{1}{2 \alpha}.   
\end{equation}

Using equation  (\ref{33.5u}) and restriction (\ref{33.2au}) we get 
\begin{equation}
H   =   \eps_0 ( 2 \alpha {\cal P})^{-1/2}, 
 \label{33.6u}
\end{equation}
where $\eps_0 = \pm 1$, and quadratic polynomial 
${\cal P} = {\cal P}(x)$  ($x = h/H$)  is defined in (\ref{33.7}).

Here we get  
 \begin{equation}
\alpha {\cal P} > 0, 
 \label{33.8u} 
\end{equation}
instead of (\ref{33.8}). 

According to eq. (\ref{33.8u}) we get in the static case
\begin{equation} 
    x_{-} < x <  x_{+} \ \ {\rm for} \  \alpha < 0,  
               \qquad \label{33.11au}
 \end{equation}    
  and    
   \begin{equation}  
     x <  x_{-} \ {\rm or} \ x > x_{+} \ \ {\rm for} \  \alpha > 0.
                 \qquad \label{33.11bu}
 \end{equation}
(The real numbers $ x_{\pm}$ are defined in (\ref{3.9}).)

We obtain that the main equations (\ref{33.8L}) and (\ref{33.M}) for  the ratio $x = h/H$
are unchanged in static case.

Thus, for $ \lambda  \neq -  21/80$ and restrictions  (\ref{33.3}), (\ref{33.2au})
imposed, we obtain  exact solutions for $H$ and $h$, which are given by 
formulae  (\ref{30x}),  (\ref{33.6u}) and (\ref{33.8u}). For $ \lambda  = -  21/80$
we should use  (\ref{32Xex}) instead of (\ref{30x}).

{\bf Graphical analysis.}
The graphical representation of $\Lambda |\alpha|$  upon $x = h/H$ in the static case
is presented at Figure 3. Here we use the relation
\begin{equation}
  \Lambda |\alpha|  = \lambda {\rm sgn}(\alpha) = 
  f(x) \frac{( 2 + 20x + 20x^2)}{|2   + 20x  + 20x^2|}, 
     \label{32Ls}
    \end{equation} 
in agreement with  (\ref{33.8u}). It follows from Figure 3 that
real solutions take place if
\begin{equation}
  \Lambda |\alpha|  \geq - \lambda_c = - 3/10 , 
     \label{32pcc}
    \end{equation} 
for $\alpha < 0$ and
\begin{equation}
  \Lambda |\alpha|  < \lambda_a = - 3/14, 
     \label{32pac}
    \end{equation}  
for $\alpha > 0$. 

In static case real solutions for $\Lambda = 0$ exists only if $\alpha <0$.

\begin{figure}[!h]
 	%\begin{minipage}[h]{0.49\linewidth}
 		\begin{center}
 			\includegraphics[width=\linewidth]{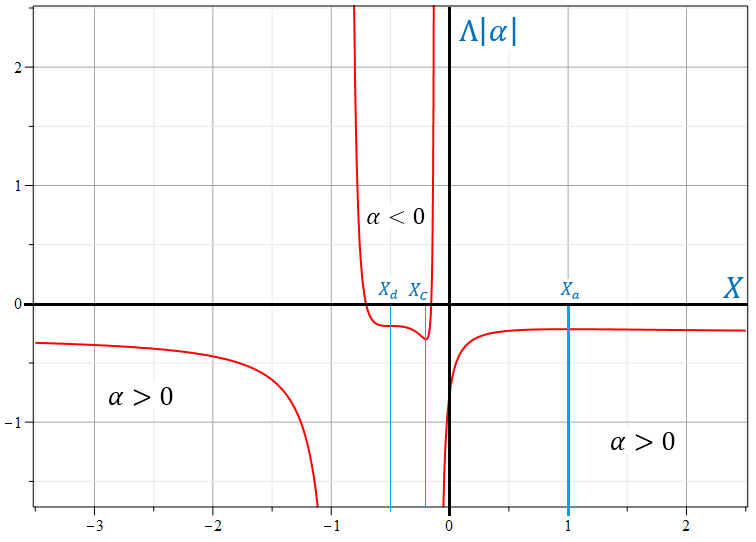}
 			\caption{The dependence of $\Lambda |\alpha|$  upon $x = h/H$ in static case.
 			The central branch 	coresponds to $\alpha <0$. The left and right branches 
 			corespond to $\alpha > 0$.}
 			\label{fig:3}
 		\end{center}
 	%\end{minipage}
  \end{figure}

{\bf Stability.} 

Using results of Ref. \cite{Ivas-20} we can analyse
the  stability of static solutions under consideration 
obeying  $x = h/H \neq x_i$, $i = a,c,d$,
where $x_a =1$,  $x_c = - \frac{1}{5}$,  $x_d = - \frac{1}{2}$. 

The solutions are  stable for $H > 0$ and $u \to + \infty$ if 
i) $x > x_d = -1/2$ and unstable if ii) $x < x_d = -1/2$.
For  $H > 0$ and $u \to - \infty$ they stable for 
i) $x < x_d = -1/2$ and unstable if ii) $x > x_d = -1/2$.

The solutions are  stable for $H < 0$ and $u \to + \infty$ if 
i) $x < x_d = -1/2$ and  unstable if ii) $x > x_d = -1/2$.
For  $H < 0$ and $u \to - \infty$ they stable for 
i) $x > x_d = -1/2$ and  unstable if ii) $x < x_d = -1/2$.

\section{Conclusions}

Here we have considered   Einstein-Gauss-Bonnet-Yang-Mills-$\Lambda$  gravitational model in dimension  $D = 10$ with  non-zero constant $\alpha$ coupled to a sum of Yang-Mills and Gauss-Bonnet terms.    

We  have studied so-called cosmological type solutions with the  metrics  (\ref{0.1})
  %$ ds^2 = - w d \chi \otimes d \chi  + a^2_3(\chi )g^{(3)} + a^2_6(\chi )g^{(6)}$ %  
defined on product manifolds $M = \R \times \R^3 \times K$, where $w = \pm 1$, 
$\R^3$ is  flat $3d$ subspace  with the metric $g^{(3)}$, and  $K$  is $6d$ 
 Ricci-flat Calabi-Yau manifold 
  with the metric $g^{(6)}$. The gauge field 1-form was  considered 
to be coinciding with  spin connection 1-form  on $K$: $A =  \omega^{(6)}$. 
       
For $w = +1$,  $\chi = t$, we have obtained exact cosmological  solutions  with exponential dependence of scale factors
(upon $t$-variable), governed by two non-coinciding Hubble-like parameters: $H >0$, $h$, 
corresponding to factor spaces of dimensions $3$  and $6$, respectively, 
when  the  following restriction: $ 3H + 6 h \neq 0$
is used (excluding the solutions with constant volume factor). 

 Static analogs of cosmological solutions ($w = - 1$, $\chi = u$ ) with exponential dependence of scale factors and 
non-coinciding ``Hubble-like'' parameters  $H \neq 0$ and $h$, obeying  $ 3H + 6 h \neq 0$, 
are also presented here. 

We have also outlined  the stability of the  solutions in cosmological case (for $t \to  + \infty$, Section 3) 
and in static case for ($u \to \pm \infty$, Section 4) and have singled out  ``islands'' of stable/non-stable solutions.

Some cosmological applications of the model ($w = 1$) may be of interest in context of
dark energy problem and  problems of stability/variation of gravitational constant. For static case ($w = -1$) possible  applications of the obtained solutions may  be a subject of a further research, aimed at a search of topological black hole solutions (with flat horizon) or wormhole solutions which are coinciding  asymptotically  (for $u \to \pm \infty$)  with our solutions.

%\begin{center}
 {\bf Acknowledgments}
%\end{center}

This research was funded by RUDN University, scientific project number FSSF-2023-0003.

 \end{document}